# Atomic-Resolution Visualization and Doping Effects of Complex Structures in Intercalated Bilayer Graphene

Jason P. Bonacum[1], Andrew O'Hara[1], De-Liang Bao[1,2], Oleg S. Ovchinnikov[1,3], Yan-Fang Zhang[1], Georgy Gordeev[4], Sonakshi Arora[4], Stephanie Reich[4], Juan-Carlos Idrobo[5], Richard F. Haglund[1], Sokrates T. Pantelides[1,5,6], Kirill Bolotin[1,4]

[1]*Department of Physics and Astronomy, Vanderbilt University, Nashville, TN, USA*
[2]*Institute of Physics and University of Chinese Academy of Sciences, Chinese Academy of Sciences, Beijing 100190, China*
[3]*Institute for Functional Imaging of Materials, Oak Ridge National Laboratory, Oak Ridge, TN, USA*
[4]*Freie Universität Berlin, Fachbereich Physik, Institut für Experimentalphysik, Arnimallee 14, 14195 Berlin, Germany*
[5]*Center for Nanophase Materials Sciences, Oak Ridge National Laboratory, Oak Ridge, TN, USA*
[6]*Department of Electrical Engineering and Computer Science, Vanderbilt University, Nashville, TN 37235, USA*

Molecules intercalating two-dimensional (2D) materials form complex structures that have been mostly characterized by spatially averaged techniques. Here we use aberration-corrected scanning transmission electron microscopy and density-functional-theory (DFT) calculations to study the atomic structure of bilayer graphene (BLG) and few-layer graphene (FLG) intercalated with $FeCl_3$. In BLG we discover two distinct intercalated structures that we identify as monolayer-$FeCl_3$ and monolayer-$FeCl_2$. The two structures are separated by atomically sharp boundaries and induce large but different free-carrier densities in the graphene layers, $7.1 \times 10^{13}$ cm$^{-2}$ and $7.8 \times 10^{13}$ cm$^{-2}$ respectively. In FLG, we observe multiple $FeCl_3$ layers stacked in a variety of possible configurations with respect to one another. Finally, we find that the microscope's electron beam can convert the $FeCl_3$ monolayer into FeOCl monolayers in a rectangular lattice. These results reveal the need for a combination of atomically-resolved microscopy, spectroscopy, and DFT calculations to identify intercalated structures and study their properties.

*This manuscript has been authored by UT-Battelle, LLC under Contract No. DE-AC05-00OR22725 with the U.S. Department of Energy. The United States Government retains and the publisher, by accepting the article for publication, acknowledges that the United States Government retains a non-exclusive, paid-up, irrevocable, worldwide license to publish or reproduce the published form of this manuscript, or allow others to do so, for United States Government purposes. The Department of Energy will provide public access to these results of federally sponsored research in accordance with the DOE Public Access Plan (*http://energy.gov/downloads/doe-public-access-plan)*.*

# I. INTRODUCTION

Graphite intercalation compounds (GICs), assemblies of foreign atoms or molecules in the van der Waals gaps between the carbon layers, have been studied for over a century for potential applications in energy storage, high-temperature superconductivity, and reaction catalysis [1–5]. The recent ability to isolate graphite with controlled number of carbon layers led to a surge of interest in intercalated few-layer graphene (FLG) and bilayer graphene (BLG) by either single atomic species, e.g. Li or Na, or molecules [6–10]. Few-layer graphene intercalated with iron chloride (FLG-$FeCl_3$) is a particularly interesting example of such a compound, although $FeCl_3$-intercalated BLG (BLG-$FeCl_3$) has received limited attention [11]. Experiments have found that the presence of $FeCl_3$ causes decoupling of the carbon layers, resulting in a graphene-like band structure and induces a very large carrier density up to $10^{14}$ $cm^{-2}$ in the graphene sheets (corresponding to a Fermi level shift as large as 1.3 eV below the Dirac point) [12]. Highly doped graphene and intercalated graphite are interesting for the study of exotic superconductivity [13,14]. In addition, it has been suggested that FLG-$FeCl_3$ develops an interesting magnetic structure with ferromagnetic order inside each $FeCl_3$ layer and antiferromagnetic coupling between the neighboring layers [15]. Such order is especially interesting in the context of the recent interest in two-dimensional magnetism [16–20]. Finally, FLG-$FeCl_3$ is stable in the ambient conditions over months, resists to degradation by common solvents, and has both high conductivity and high optical transparency [11,12]. These properties invite potential applications for energy storage, transparent conductors, and heat spreaders [21,22].

At the same time, while multiple experiments addressed the macroscopic properties of FLG-$FeCl_3$, its microscopic structure remains virtually unknown and the possibility that multiple intercalant structures form has not been adequately explored. Electron diffraction and powder X-ray diffraction (XRD) data from $FeCl_3$-GICs suggest that within each van der Waals gap the $FeCl_3$ molecules form a honeycomb lattice similar to bulk $FeCl_3$ [2,23]. XRD, however, being a spatially averaged technique, is not sensitive to several possibilities. For example, intercalant layers in FLG may have layer-number dependent properties, as is known to occur in the lithium intercalation process [24]. Such properties should be studied to understand the predicted antiferromagnetic coupling between neighboring $FeCl_3$ layers [15]. Previous research also demonstrates that $FeCl_3$ is converted to $FeCl_2$ in a reducing environment, but the published work on the stability of FLG-$FeCl_3$ does not consider the possibility of $FeCl_2$ formation [2]. Lattice defects in FLG-$FeCl_3$, if present, have not so far been investigated by any means, but are expected to strongly scatter the charge carriers in graphene layers, thereby limiting applicability of FLG-$FeCl_3$ in electronics. The presence of defects should also affect the dynamics of the intercalation process and thus be critical for energy storage applications.

In this paper, we report atomic-resolution structures of FLG and BLG intercalated with $FeCl_3$. Aberration-corrected scanning transmission electron microscopy (STEM) of BLG-$FeCl_3$ reveals two distinct structures that, in combination with density-functional-theory (DFT) calculations and STEM image simulations, are identified as monolayer-$FeCl_3$ and monolayer-$FeCl_2$. Both structures exhibit a hexagonal crystal structure, sharp boundaries between intercalated and unintercalated regions, and lattice defects. The presence of two structures and their effects on doping are confirmed via electron energy loss spectroscopy (EELS) and Raman spectroscopy and discussed in the context of DFT calculations. We also observe a rectangular lattice of FeOCl that is formed at the edges of the $FeCl_3$ monolayer when exposed to the electron beam during STEM imaging. In $FeCl_3$-intercalated FLG, we observe stacked $FeCl_3$ monolayers that display no preferred orientation of the $FeCl_3$ from layer to layer. Our results shed light on the origin of doping in intercalated graphene, its stability, and may help in understanding magnetic properties of intercalated systems.

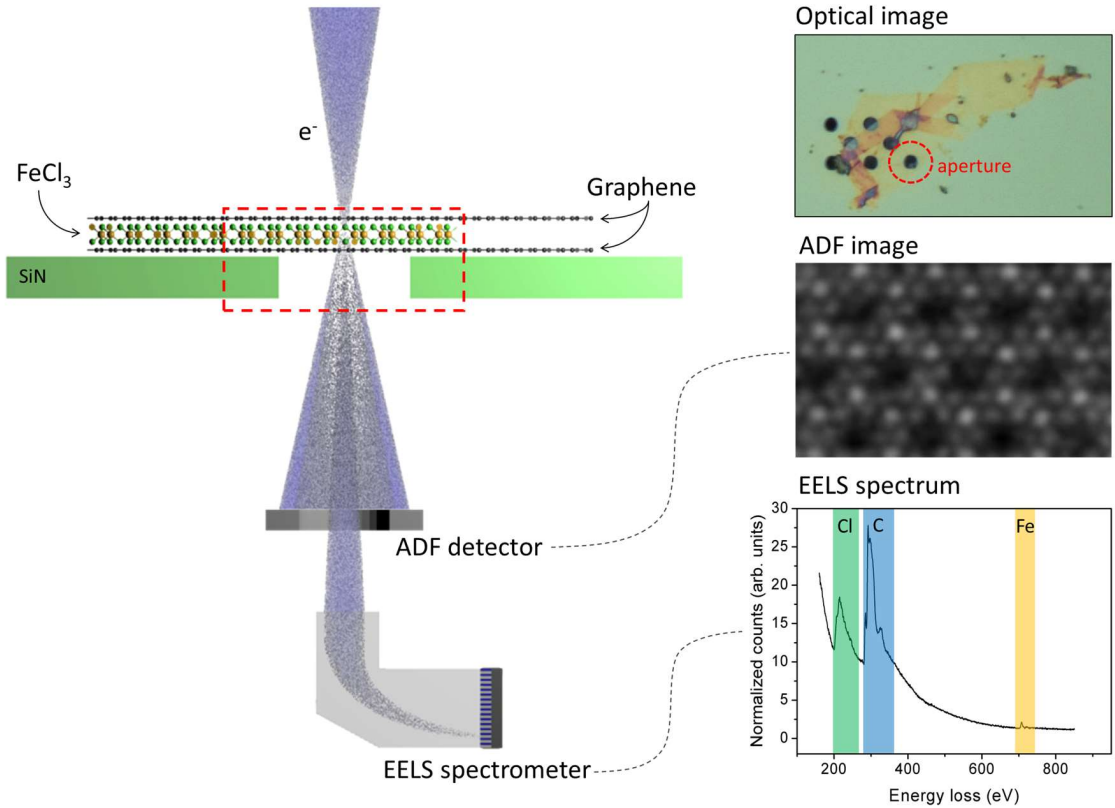

FIG. 1. (Left) Diagram of FeCl$_3$-BLG inside a STEM with EELS capabilities. (Top right) Optical image of the same sample with a dotted red outline showing the region where STEM is performed. (Center right) ADF image of the same sample. (Bottom right) EELS of the same sample with labels on the signals for chlorine (green), carbon (blue), and iron (yellow) atoms.

The samples were fabricated using the vapor transport method of intercalation [2,12]. Bilayer graphene and few-layer graphene were transferred onto holes (2 μm diameter) in silicon nitride membranes, as seen in the top right panel of Fig. 1. The samples were vacuum sealed in borosilicate ampules with anhydrous FeCl$_3$ and then transferred to a tube furnace for the intercalation reaction. After intercalation, the samples were washed with deionized water to remove any adsorbed FeCl$_3$ that could interfere with imaging of the intercalated FeCl$_3$. Raman spectroscopy was performed before and after intercalation to confirm the presence of FeCl$_3$, which is evidenced by new blue-shifted G peaks [12]. Additional experimental details are provided in the Methods section and in the Supplemental Material.

## II. ATOMIC STRUCTURE OF FeCl$_3$ MONOLAYER

Aberration-corrected STEM was used to investigate the atomic structure of the resulting intercalants. Atomic-resolution images were obtained using an annular dark field (ADF) detector, and the elemental composition was confirmed using EELS, as illustrated in Fig. 1, where we show data for a FeCl$_3$-BLG sample. While intercalated iron and chlorine are clearly resolved, carbon atoms are barely visible as the ADF signal strength is roughly proportional to the square of the atomic number. Thus, the Fe and Cl signals are roughly nineteen and eight times stronger, respectively, than the carbon signals. A key result of this investigation is that we were able to identify both FeCl$_3$ and FeCl$_2$ in adjacent regions, which indicates that FeCl$_3$ molecules can undergo reduction within the van der Waals gap of BLG. We first present the data and analysis for FeCl$_3$ intercalants. The pertinent ADF image [Fig. 2(a)] exhibits a 2D honeycomb structure. This structure is the same as in bulk FeCl$_3$ with each iron atom bonded to six chlorine atoms in an octahedral geometry, as shown in Fig. 2(b). The carbon atoms can be seen faintly inside the holes of the FeCl$_3$ honeycomb lattice displayed in Fig. 2(a), although the contrast is much lower than that of the chlorine and iron atoms. In areas where there is incomplete intercalation [Fig. 2(d)], the intercalants form islands that are separated from neighboring unintercalated regions by an atomically sharp boundary. This sharpness is due to in-plane covalent bonds between the iron chloride molecules in the intercalant layer.

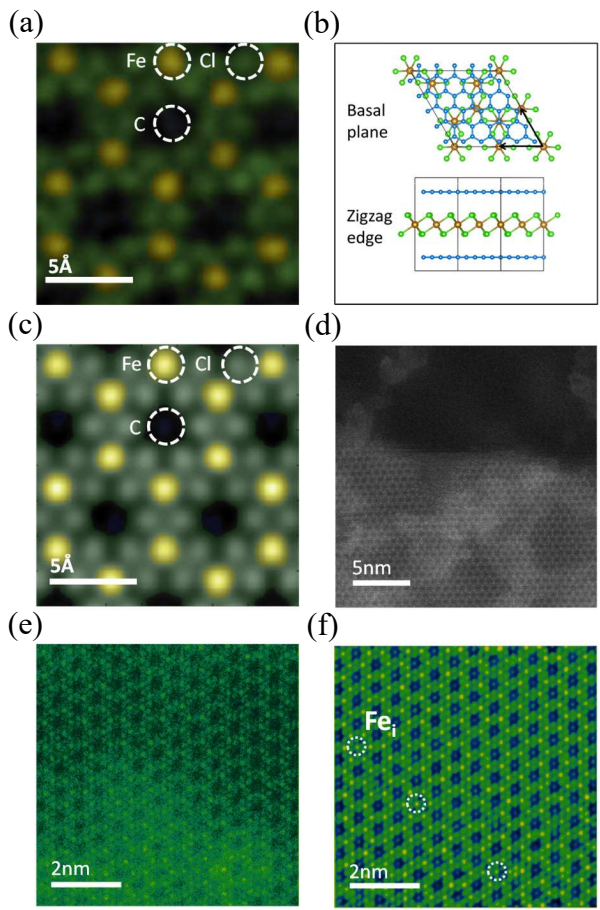

FIG. 2. (a) Colorized ADF image of FeCl$_3$-BLG. (b) Diagram of FeCl$_3$-BLG with black arrows displaying the primitive lattice vectors of FeCl$_3$. (c) Colorized STEM simulation of FeCl$_3$-BLG. (d) ADF image of intercalation boundary. (e) Unfiltered ADF image of FeCl$_3$-BLG used for PCA filtering. (f) Same image as in panel (e), but filtered using components 2-10 of the PCA. Iron interstitial defects are highlighted by white dotted outlines as visible (the colors are determined by a color scale ranging from dark blue to yellow).

To unambiguously determine the position of each atomic species with STEM simulations, atomic positions optimized by DFT calculations and the experimental beam parameters were used as inputs to the QSTEM software package to create simulated ADF images [Fig. 2(c)] for comparison to the experimentally obtained images [25]. There is good agreement between the ADF image and STEM simulation. The measured FeCl$_3$ lattice constant of 0.61±0.01 nm agrees with the theoretical value of 0.60 nm. These data demonstrate that the intercalated FeCl$_3$ forms a 2D material in between the graphene layers.

In order to better resolve carbon atoms and reduce surface contamination effects, we filtered the original data [Fig. 2(e)] using principal component analysis (PCA) [26,27]. We noticed that the first PCA component corresponds primarily to the surface contamination and the higher order components (>10) correspond to background noise in the image. We therefore plotted the components 2 through 10 [Fig. 2(f)]. The three types of atoms present in the samples are visible in the images. While the background is dark blue, iron atoms appear yellow, chlorine atoms – green, and carbon atoms – light blue. These light blue spots are separated by 0.57±0.01 nm, which corresponds to four times the carbon-carbon bond in graphene, and sometimes appear off center or as dumbbells inside the hexagons of the FeCl$_3$ lattice. These observations further confirm the source of the light blue spots as the carbon lattice and not an artifact from the FeCl$_3$ structure, which would have similar hexagonal symmetry. The location of the carbon atoms indicates that the carbon lattice and FeCl$_3$ lattice are aligned with each other in this sample. The PCA filtering also displays interstitial iron atoms in the FeCl$_3$ hexagons. While interstitial iron can be seen on the left-center edge of the unfiltered image [Fig. 2(e)], the removal of the surface contamination in the image makes it clear that several such interstitials occur in this section of the sample. Such additional interstitial iron atoms at non-regular lattice sites are likely to impact the magnetic ordering properties of FeCl$_3$-intercalated FLG [15].

## III. ATOMIC STRUCTURE OF FeCl$_2$ MONOLAYER

We now turn to the region of the intercalant structure that is different from the honeycomb structure described above, which we identify as FeCl$_2$. The pertinent ADF image is shown in Fig. 3(a). The structure in Fig. 3(a) looks similar to a monolayer of FeCl$_3$, but with an additional iron atom in the holes of the honeycomb lattice, highlighted yellow in Fig. 1. These features include a step-like edge and Lorentzian-shaped peaks, referred to as white-lines [Fig. 3(c)]. The ratio of the L$_3$ and L$_2$ white-line intensities is used to differentiate between Fe$^{2+}$ and Fe$^{3+}$ species, which have L$_3$/L$_2$ intensity ratios of 4.0 and 5.5 respectively [28]. Experimentally, we determined that ratio using both Lorentzian fits as well the analysis of the second derivative of the data (see Supplemental Material for more details) [28]. Both methods yield the ratio of 4±1, consistent with the structure being FeCl$_2$. The literature on intercalation of bulk graphite with FeCl$_3$ demonstrates that FeCl$_2$ cannot be intercalated on its own, but the FeCl$_3$ can be reduced to FeCl$_2$ after intercalation into the graphite [2]. Given this evidence in the literature, we suggest that the FeCl$_3$ is reduced to FeCl$_2$ after the intercalation reaction. This happens due to the presence of a reducing agent, which could be hydrocarbon contaminants and hydrogen formed from other chemical reactions during the intercalation process.

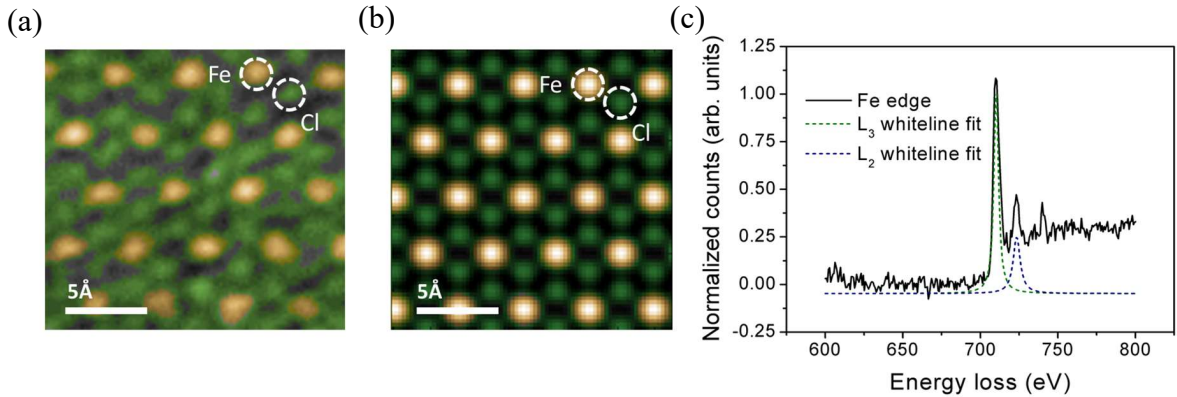

FIG. 3. Colorized (a) ADF image and (b) STEM simulation of FeCl$_2$ in BLG. (c) EELS spectra of the area shown in panel (a) with Lorentzian fits of the L$_3$ and L$_2$ white lines.

signifying a change in stoichiometry from FeCl$_3$ to FeCl$_2$. Comparison of this structure to a STEM simulation of monolayer FeCl$_2$ [Fig. 3(b)] exhibits good agreement, with an experimental lattice parameter of 0.350±0.005 nm compared to the theoretical value of 0.347 nm. We further probed the same region by EELS, which is sensitive to the iron oxidation state. We focus on spectral features corresponding to iron core electron excitations,

## IV. DOPING EFFECTS OF FeCl$_3$ AND FeCl$_2$

### A. Raman spectroscopy

We find further evidence for the coexistence of FeCl$_3$ and FeCl$_2$ regions in FeCl$_3$-BLG in their effects on the free carrier density of graphene. The carrier density in graphene, as well as its proxy, the position of the Fermi energy relative to the Dirac point, were probed via Raman spectroscopy. Before intercalation, the Raman G mode peak is at 1582 cm$^{-1}$. After intercalation, the G peak splits into three peaks – G0, G1, and G2 at 1586 cm$^{-1}$, 1614 cm$^{-1}$, and 1626 cm$^{-1}$ respectively [Fig. 4(a)]. Since the spectral position of the G mode is indicative of the local free carrier density of graphene, such splitting is consistent with the presence of regions with three distinct carrier densities within the diffraction-limited laser spot on the sample [29,30]. With its spectral position within 4 cm$^{-1}$ from the G peak before intercalation, the G0 peak indicates undoped graphene, while G1 and G2 correspond to higher carrier densities. To determine these carrier densities quantitatively, we varied the laser excitation energy. The peaks G1 and G2 exhibit maximum intensities at 1.96 eV and 2.07 eV excitation energies respectively, while the G0 peak intensity is relatively constant with excitation energy [Fig. 4(b)]. The maximum in G peak intensity at a certain excitation energy signals that the local Fermi energy is half the excitation energy [31]. Assuming that FeCl$_3$ is an acceptor molecule, we therefore determine that the Fermi energy corresponding to G0 is at the local Dirac point, while those for G1 and G2 are 0.98 eV and 1.03 eV below the local Dirac point, respectively, as illustrated in Fig. 4(c). As the graphene layers are essentially decoupled, the number of free carriers in each region can be approximated via the relation $n \approx \frac{1}{\pi}(E_F/v_F \hbar)^2$. We determine the carrier densities of ~0 cm$^{-2}$, $7.1 \times 10^{13}$ cm$^{-2}$ and $7.8 \times 10^{13}$ cm$^{-2}$ for G0, G1, and G2 respectively.

## B. Discussion

Combining the Raman spectroscopy results with the STEM data, we draw several conclusions about the atomic origin of the different free carrier densities. In the literature, the appearance of two blue-shifted G peaks is attributed to staging or surface adsorption of FeCl$_3$ [12,32]. However, staging does not occur in BLG. In our STEM images, there is only a single monolayer of FeCl$_3$ between the layers of graphene and no adsorbed FeCl$_3$ on its surface. The presence of both the G1 and G2 peaks therefore evinces the presence of two different types of intercalants that locally induce different doping levels. We hypothesize that the coexistence of FeCl$_3$ and FeCl$_2$, as portrayed in Fig. 4(c), is responsible for the two positions of the universal Fermi energy relative to the local Dirac points. This hypothesis is supported by DFT calculations that exhibit two different positions for the Fermi level below the Dirac point, 0.42 eV for FeCl$_2$ and 0.66 eV for FeCl$_3$. The smaller theoretical values of the Fermi energy relative to the local Dirac point compared to the experimental values is likely due to the underestimation of the Fermi velocity within the local density approximation [33].

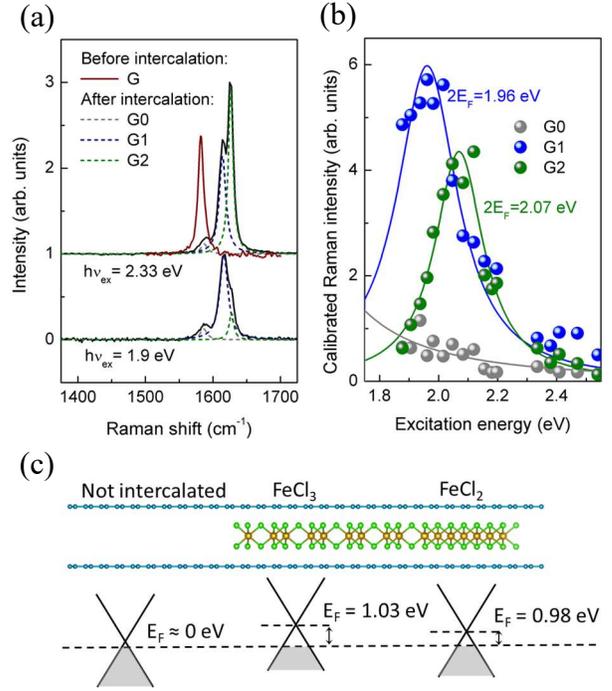

FIG. 4. (a) Raman spectra for intercalated BLG-FeCl$_3$ in the region of the G band with 1.9 eV (bottom) and 2.33 eV (top) excitation energies shown in black. In addition, the G band of pristine BLG before intercalation is shown in the top graph in red. The dotted lines are Lorentzian fits of the G0, G1, and G2 peaks. (b) Raman intensity of the G bands as a function of excitation energy. The peak maximum is achieved when the laser energy matches twice the Fermi energy. (c) A cartoon depiction of BLG intercalated with both FeCl$_3$ and FeCl$_2$ above a diagram of the respective relative Fermi energies.

The appearance of free carriers in graphene layers adjacent to another material is typically interpreted as charge transfer. We used the present DFT-calculated charge densities in the intercalated BLG to test this interpretation. We found that, while the Fermi energy deviates from the Dirac point in $FeCl_3$-BLG, there is virtually no net charge transfer between the graphene and intercalant layers. Although the drop of the Fermi level below the Dirac point suggests a net transfer of electrons from the graphene to the intercalant, the wavefunctions from the valence states in the $FeCl_3$ extend into the graphene layers thereby maintaining overall charge neutrality. In other words, the proximity of graphene to another material causes a redistribution of electrons in the *energy space* to produce free carriers (doping), seemingly corresponding to charge transfer, the distribution of electrons in physical space actually remains relatively unchanged (see Supplemental Material for details).

Overall, our DFT and Raman spectroscopy data suggest the association of the peaks G0, G1, and G2 with unintercalated regions, regions intercalated with $FeCl_2$, and regions intercalated with $FeCl_3$ respectively. Our STEM data are consistent with this assignment. The presence of the G0 peak is corroborated by the observation of unintercalated regions in STEM images.

## V. ALIGNMENT OF MULTIPLE $FeCl_3$ MONOLAYERS

### A. Interpreting the alignment from atomic-resolution images

We also imaged intercalated FLG with thicknesses of four to six graphene layers to study the relative angular alignment of $FeCl_3$ monolayers sandwiched between successive layers of graphene and test whether a superposition of $FeCl_3$ and $FeCl_2$ layers needs to be invoked to reproduce the images. The ADF images [Fig. 5] reveal only $FeCl_3$ layers with different degrees of alignment for different samples. The first sample [Fig. 5(a)] exhibits complete angular alignment of the $FeCl_3$ layers, observed as coincident ADF signal from the atoms in each layer. The stacking configuration is inferred by comparing the ADF image with STEM simulations of bilayer and trilayer $FeCl_3$ and $FeCl_2$ in different stacking configurations. Two and three layers were used since the sample had approximately four to six layers of graphene, determined by atomic force microscopy, and the Raman spectrum of the sample after intercalation indicates partial intercalation (see Supplemental Material for more details). The best agreement between the ADF image and STEM simulation is for ABC stacking of $FeCl_3$ when comparing AA, AB, AAA, ABA, and ABC stacking configurations of both $FeCl_3$ and $FeCl_2$. The STEM simulation for ABC-stacked $FeCl_3$ is shown in Fig. 5(b), while the simulations for the other stacking configurations can be found in the Supplemental Material.

The second sample exhibits small angles of rotation between the intercalant layers, producing a moiré pattern in the ADF image [Fig. 5(c)]. The relative angles between each monolayer were determined from the fast Fourier transformation (FFT) of the ADF image. The FFT of the second sample [Fig. 5(d)] displays sharp Fourier peaks in a hexagonal pattern due to the hexagonal structure of the crystal basis for $FeCl_3$. We assume only $FeCl_3$ is present in this section of the sample given the greater relative abundance of $FeCl_3$ compared $FeCl_2$ in the previous samples. Each layer of $FeCl_3$ has a distinctive set of Fourier peaks, and the relative angle between the layers can be observed from the angles between those Fourier peaks. Three distinct sets of peaks can be seen in Fig. 5(d) with angles of 0º, 3.0º, and 5.5º. Given these angles, the lattice parameter of the moiré pattern ($a_{moiré}$) can be calculated using the following equation (see Supplemental Material for further details):

$$a_{moiré} = \frac{a}{\sqrt{2(1-\cos\theta)}} \quad (1)$$

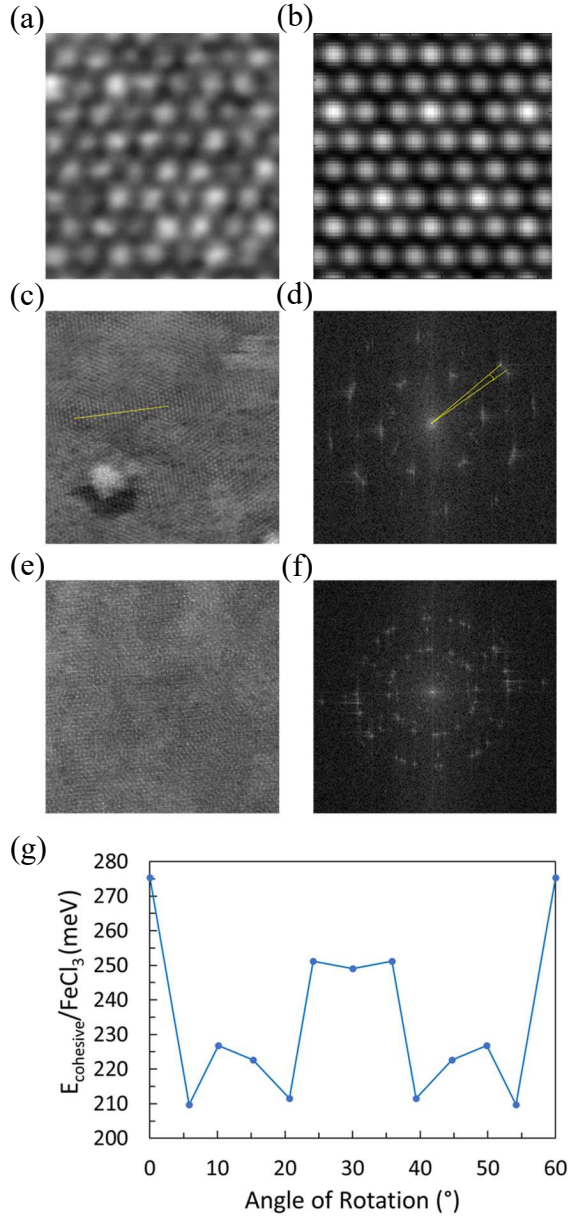

FIG. 5. (a) Annular dark-field (ADF) image of sample 1, suggested structure: ABC-stacked $FeCl_3$. (b) STEM simulation of ABC-stacked $FeCl_3$. (c) ADF image of sample 2, suggested structure: nearly aligned ABC-stacked $FeCl_3$. (d) Fast fourier transform (FFT) of the ADF image for sample 2. (e) ADF image of sample 3, suggested structure: $FeCl_3$ with uncorrelated stacking. (d) FFT of the ADF image for sample 3. (g) Cohesive energy per $FeCl_3$ molecule inside BLG vs. relative orientation between the crystalline axes of graphene and $FeCl_3$ obtained from DFT calculations.

where $a$ is the lattice parameter of the $FeCl_3$ measured from the ADF image and $\theta$ is the relative angle between each layer measured from the FFT. Using the values of 3.0° and 2.5° in equation 1, the moiré lattice parameter is 12nm and 14nm respectively. The moiré patterns for these angles cannot be seen in the ADF image [Fig. 5(c)] as they are too large for the size of the image, but 5.5° gives a moiré lattice parameter of 6.3 nm, which agrees with the moiré lattice parameter, 6.2±0.1 nm, seen in Fig. 5(c).

The third sample exhibits angles between $FeCl_3$ layers as large as 44° measured in the FFT [Fig. 5(f)], that corresponds to $a_{moire}$ of 0.81 nm, close to the lattice parameter of $FeCl_3$ (0.607 nm). There are also at least six distinct Fourier peaks spread across the 44° creating an FFT that resembles that of a polycrystal. The ADF image for this sample [Fig. 5(e)] appears disordered due to the number of layers and wide range of angles, but faint moiré patterns can still be seen from the layers of $FeCl_3$ that have small relative angles to each other, which correspond to moiré-pattern lattice parameters on the order of nanometers.

## B. Explaining the results with DFT calculations

To gain insight into the source of the observed moiré patterns and apparent polycrystalline structure, we calculated the interlayer cohesion energetics as a function of twist angle between $FeCl_3$ and graphene, as shown in Fig. 4(g). For the calculated angles, we find that there is a global energy minimum at 0°/60° and additional local energy minima at 10°, 25°, 35°, and 50°. However, the overall range of energies is only 65 meV per $FeCl_3$ unit. That energy range is significantly smaller than the available thermal energy (~6kT = 300 meV per $FeCl_3$ unit) suggesting that patches can nucleate with essentially any relative angle. Although bulk $FeCl_3$ orders with a relative AB stacking between layers, the presence of a graphene layer between two layers of $FeCl_3$ renders the two $FeCl_3$ stacking configurations degenerate. This result implies that in addition to the presence of relative twist angles between intercalant layers,

## VI. FORMATION OF FeOCl

Finally, we observe the formation of a new monolayer that appears at the edge of the intercalated FeCl$_3$ monolayers in FeCl$_3$-BLG when exposed to the electron beam under STEM imaging conditions with 60 kV accelerating voltage [Fig. 6(a)]. Such monolayers are never seen at the beginning of STEM imaging and cannot affect the interpretation of our Raman results that are recorded before STEM. This new monolayer forms a rectangular lattice composed of iron, chlorine, and oxygen as shown by EELS in Fig. 6(b). The constituent components and lattice shape suggest that the compound is iron oxychloride (FeOCl), a material that has previously been described in bulk layered form [33,34]. The identification of the material is corroborated in Fig. 6(c) by an overlay of the FeOCl monolayer atomic structure, the experimental ADF image, and a STEM simulation. DFT calculations of the monolayer FeOCl electronic structure [Fig. 6(d)] indicate that the ferromagnetic and antiferromagnetic states are almost degenerate with indirect band gaps of 2.70 eV and 2.50 eV, respectively. Further exploration of the properties of this monolayer is deferred to a future paper. The creation of this novel monolayer in the FeCl$_3$-intercalated bilayer system suggests the ability to engineer additional interesting materials and structures after the initial synthesis. More specifically, the microscope's electron beam

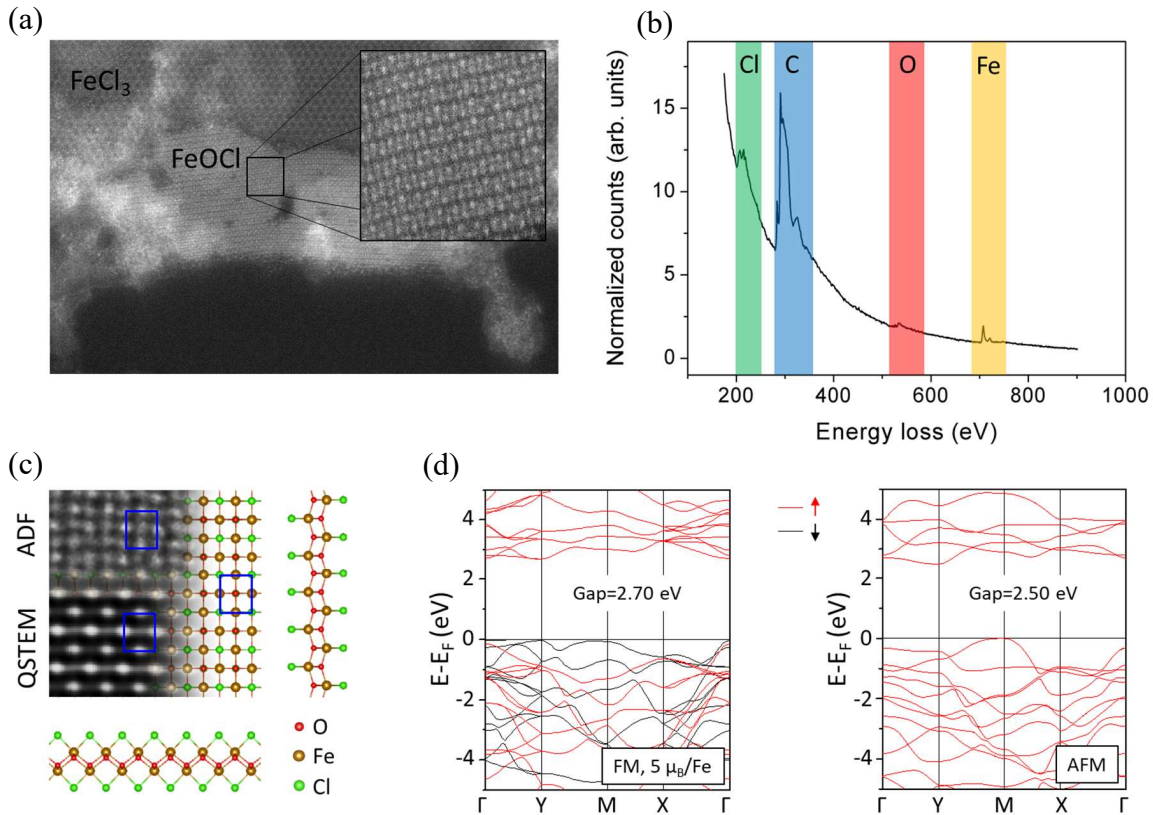

FIG. 6. (a) ADF image of the edge of the FeCl$_3$ monolayer after it has been irradiated during imaging. The new rectangular structure that is visible is interpreted as FeOCl. (b) EELS of the same region. (c) ADF image along with the STEM simulation of FeOCl. (d) Calculated band structure of FeOCl for ferromagnetic (left) and antiferromagnetic (right) ordering.

is in effect used to "process" intercalants and convert them into new structures.

## VII. CONCLUSIONS

In conclusion, this work demonstrates intercalation of molecules in BLG or FLG can lead to the formation of diverse complex structures. We observed the formation of crystalline FeCl$_3$ monolayers with a honeycomb structure like that of bulk FeCl$_3$, atomically sharp boundaries between intercalated and unintercalated regions, the presence of defects, and a variety of possible orientations for the FeCl$_3$ relative to graphene layers. This information is useful for the study of interesting phenomena in graphene such as Klein tunneling of Dirac-like fermions, which requires atomically sharp doping boundaries, and the study of effects on the electronic band structure of graphene due to superlattice formation between the FeCl$_3$ and graphene honeycomb lattices [36,37]. The observation of iron interstitial defects also has possible applications in information storage due to modification of the local magnetic field by the defects. We provide evidence for coexistence of both FeCl$_3$ and FeCl$_2$ in BLG under ambient conditions, which was not observed previously and could provide a new perspective for interpreting the stability of FeCl$_3$-intercalated FLG. Specifically, our results suggest that previously reported changes in the Raman spectra of FLG-FeCl$_3$ might be due to the formation of FeCl$_2$ rather than deintercalation of FeCl$_3$ [11,12]. Additionally, we demonstrate the conversion of monolayer FeCl$_3$ into FeOCl via an electron-beam-induced reaction inside BLG, revealing intercalated BLG to be a useful vessel for creating novel 2D materials.

## METHODS

### A. Few-layer graphene (FLG) fabrication

The FLG was mechanically exfoliated from kish graphite onto polydimethylsiloxane (PDMS) using Scotch tape. The support grid for the FLG was prepared by milling 2 μm apertures in 50 nm silicon nitride membranes (PELCO Silicon Nitride Support Films) with a Helios Nanolab G3 CX dual beam focus ion beam – scanning electron microscope, and the FLG was then transferred to the silicon nitride membrane over the apertures using a viscoelastic stamp transfer method [38]. The BLG samples were purchased commercially (Graphene on PELCO Holey Silicon Nitride). Graphene in these samples was grown by chemical vapor deposition.

### B. Vapor transport method

The vapor transport method of intercalation involves vacuum-sealing the FLG and FeCl$_3$ powder inside an ampule and then annealing, which causes the FeCl$_3$ to evaporate and spontaneously intercalate into the FLG. The ampule is prepared by sealing one end of a ¼ inch diameter borosilicate tube with a butane torch and then baking overnight at 150 °C to remove moisture. Then 0.02 g of FeCl$_3$ were transferred to the ampule, and the ampule was evacuated to 5 mTorr with an Edwards 5 two-stage rotary-vane vacuum pump. The ampule is attached to the vacuum setup using a quick-connect coupler. To ensure that the FeCl$_3$ is anhydrous, the ampule is heated to 120 °C for 30 min during evacuation and purged three times with nitrogen gas. The sample was then inserted into the ampule, and the evacuation procedure was repeated. Once the ampule pressure gets down to 0.5 mTorr, a butane torch was used to seal the ampule approximately 10cm from the opposite end of the FeCl$_3$ powder. The ampule was then annealed is a Lindenburg Blue M 1-inch tube furnace to initiate the intercalation reaction. For the reaction process, the tube furnace was heated to 340 °C (measured at the center) with a ramp rate of 1 °C/s and PID setting of 20-120-30. The reaction takes place over six hours with the ampule 5cm from the center, which results in a temperature difference of ~15 °C between the FLG and FeCl$_3$ powder. Finally, the tube furnace was cooled at a rate of 1 °C/s, and the intercalated sample was then removed by scoring and breaking open the ampule.

### C. STEM parameters

All ADF images were acquired using an aberration-corrected Nion UltraSTEM 100™ operated at 60 kV accelerating voltage [39]. We used a semi angle convergence of 30 mrad and detection ADF semi angle range of 86-200 mrad for the intercalated FLG samples and a detection semi angle of 54-200 mrad for the intercalated BLG samples. Additionally, the length scales in our data were compared with a reference sample to ensure accuracy of the measured lengths for this work.

### D. Resonance-Raman spectroscopy

The resonance-Raman spectra were obtained at the same spot on the intercalated sample. A tunable laser system with a dye laser (Radiant dye: 550-675 nm) and an Ar-Kr laser (Coherent Innova 70c: 450-530 nm) were used to excite the sample. The laser power was limited to 500 μW to avoid heat-induced effects (x100 microscope objective). The light was dispersed by a T64000 Horba Jobin Yvon spectrometer equipped with 900 grooves per mm grating and a silicon charge-coupled device in single detection mode and backscattering configuration. Elastically scattered light was rejected by a long pass filter. Raman shift was calibrated on the benzonitrile reference molecule and the Raman intensity of pristine bilayer graphene to account for the wavelength dependent spectrometer sensitivity and interference with the substrate.

### E. Computation Details

Spin-polarized DFT calculations used the Vienna *ab initio* Simulation Package (VASP) [40] within the Perdew-Burke-Ernzerhof (PBE) form of the generalized gradient approximation [41] along with Grimme's D2 van der Waals correction [42]. Interactions between valence and core electrons were described using the projector augmented wave method [43,44] with a plane wave basis cutoff of 400 eV. For the $FeCl_3$ and $FeCl_2$ primitive cells, the Brillouin zones were sampled with Γ-centered k-point grids of $8 \times 8 \times 1$ and $20 \times 20 \times 1$, respectively. A vacuum layer of at least 15 Å was used in all calculations and interatomic forces were minimized to be less than 0.01 eV/Å. Spin-polarized band structure calculations for FeOCl were performed using the HSE06 hybrid functional [45,46].

## ACKNOWLEDGEMENTS

This work was supported by National Science Foundation Grant DMR-1508433, Department of Energy Grant DE-FG02-09ER46554, the European Research Council Starting grant 639739, and by the McMinn and Stevenson endowment at Vanderbilt University. Supercomputer time was provided by the Extreme Science and Engineering Discovery Environment (XSEDE), which is supported by National Science Foundation Grant ACI-1053575 as well as by the Department of Defense's High-Performance Computing Modernization Program (HPCMP). Sample fabrication and characterization was conducted at the Vanderbilt Institute of Nanoscale Science and Engineering (VINSE). The STEM experiments of this research were conducted at the Center for Nanophase Materials Sciences, which is a DOE Office of Science User Facility (JCI).

The authors also wish to thank Ross Koby and Adam Cohn for assisting the fabrication of samples used in this work.

## REFERENECES

Supplementary Material for

# Atomic-Resolution Visualization and Doping Effects of Complex Structures in Intercalated Bilayer Graphene

Jason P. Bonacum[1], Andrew O'Hara[1], De-Liang Bao[1,2], Oleg S. Ovchinnikov[1,3], Yan-Fang Zhang[1], Georgy Gordeev[4], Sonakshi Arora[4], Stephanie Reich[4], Juan-Carlos Idrobo[5], Richard F. Haglund[1], Sokrates T. Pantelides[1,5,6], Kirill Bolotin[1,4]

[1]*Department of Physics and Astronomy, Vanderbilt University, Nashville, TN, USA*
[2]*Institute of Physics and University of Chinese Academy of Sciences, Chinese Academy of Sciences, Beijing 100190, China*
[3]*Institute for Functional Imaging of Materials, Oak Ridge National Laboratory, Oak Ridge, TN, USA*
[4]*Freie Universität Berlin, Fachbereich Physik, Institut für Experimentalphysik, Arnimallee 14, 14195 Berlin, Germany*
[5]*Center for Nanophase Materials Sciences, Oak Ridge National Laboratory, Oak Ridge, TN, USA*
[6]*Department of Electrical Engineering and Computer Science, Vanderbilt University, Nashville, TN 37235, USA*

## I. SAMPLE CHARACTERIZATION

Fig. S1 shows the reaction vessel and an optical image of the FLG sample as described in the methods section of the main text. The $FeCl_3$-intercalated FLG samples were characterized by atomic force microscopy (AFM), Raman spectroscopy, and electron energy loss spectroscopy (EELS). Examples of these data are given in Fig. S1 and S2. Fig. S1 shows data for the FLG with nearly aligned $FeCl_3$ [STEM data in Fig. 3(c) of the main text], and Fig. S2 shows data for the intercalated BLG [STEM data is Fig. 2 of the main text]. The shape of the 2D peak around $2700 cm^{-1}$ [Fig. S1] shows that the FLG graphene is more than three layers thick, while the AFM height profile [Fig. S1] confirms a thickness of four to six layers [1,2]. When the FLG is intercalated by $FeCl_3$, the graphene layers undergo hole doping, and each graphene layer can be doped by either one $FeCl_3$ layer between the graphene layers or by two $FeCl_3$ layers on each side. This discrete doping is manifested in the Raman spectra as the appearance of two blue-shifted G peaks, because the G peak blue shifts linearly with doping due to the Kohn anomaly at the K point in the graphene phonon band structure [3–5]. For the FLG samples in this work, the Raman spectrum after intercalation indicates partial intercalation, as evidenced by blue shifting and splitting of the G peak, and the EELS of the FLG after intercalation verifies the presence of carbon, iron, and chlorine atoms.

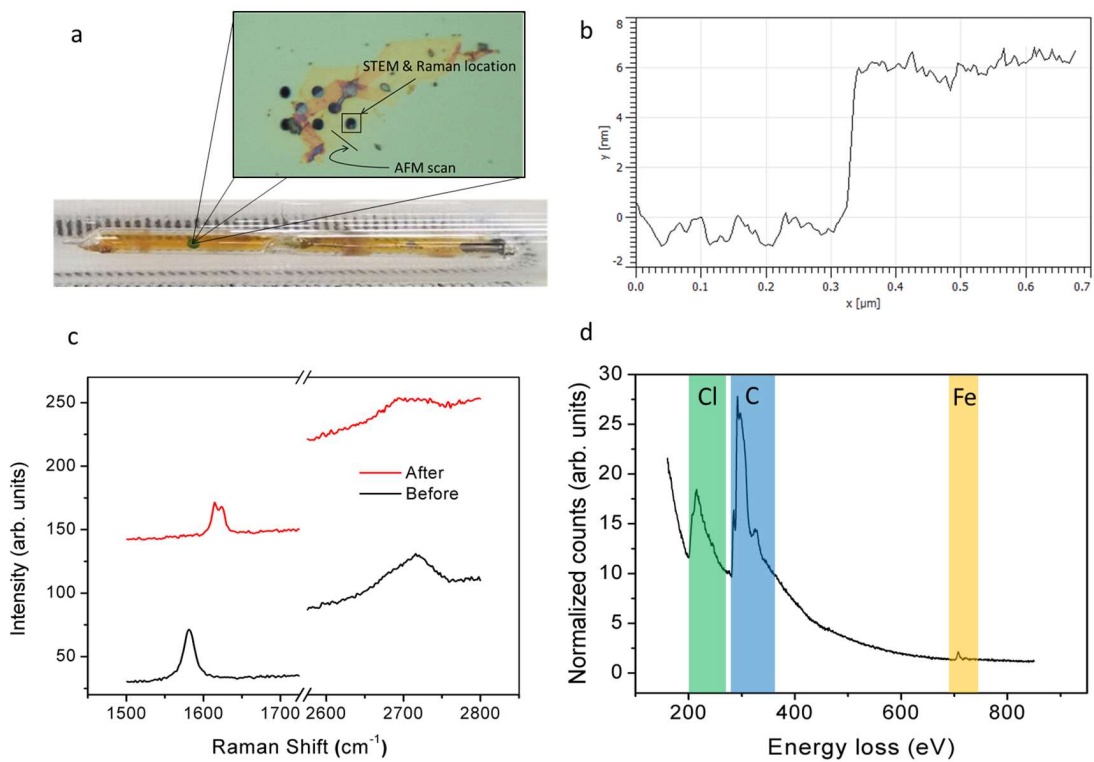

FIG. S1: (a) Image of an FeCl$_3$-intercalated FLG sample on a milled silicon nitride membrane, showing the location of STEM, Raman spectroscopy, and atomic force microscopy (AFM) measurements. Below displays an image of the vacuum-sealed reaction vessel with the sample on the left and FeCl$_3$ powder on the right. (b) AFM height profile before intercalation. (c) Raman spectra before and after intercalation (d) Electron energy loss spectrum (EELS) with labels for the chlorine, carbon, and iron signatures

For the BLG samples, the G peak may also split after intercalation, but the lower wave number peak corresponds to undoped graphene [Fig. S2]. This splitting is due to incomplete formation of the FeCl$_3$ layer inside the BLG, which is why intercalation boundaries can be seen in the STEM data [see Fig. 2(c) in the main text]. The EELS also verifies the presence of carbon, iron, and chlorine.

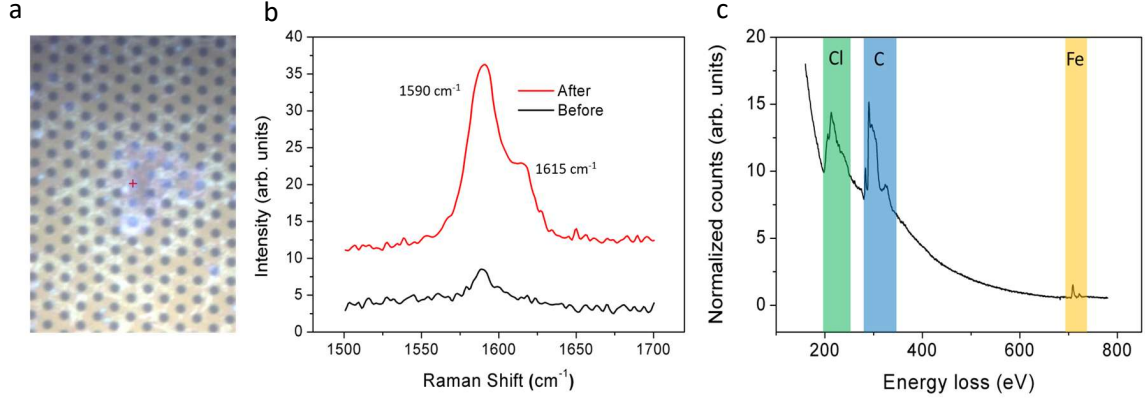

FIG. S2: (a) Image of intercalated BLG on holey silicon nitride (b) Raman spectra before and after intercalation (c) EELS with labels for the chlorine, carbon, and iron signatures

## II. UNIT CELL MATCHING FOR FeCl$_3$ / GRAPHENE

In order to calculate the rotational dependence of the binding energy between FeCl$_3$ and graphene, a commensurate unit cell with minimal strain must be determined. Within our DFT calculations, the optimal lattice parameters were 6.074 Å for FeCl$_3$ and 2.464 Å for graphene. Supercells of FeCl$_3$ were constructed in the usual manner (*i.e.* multiples of the two primitive lattice vectors). For graphene, supercells were constructed as a linear combination of lattice vectors:

$$a'_1 = n\vec{a}_1 + m\vec{a}_2 \qquad (1)$$
$$a'_2 = -m\vec{a}_1 + (n-m)\vec{a}_2 \qquad (2)$$

which gives this new supercell a twist angle $\cos^{-1}[(a'_1 \cdot a_1)/(|a'_1||a_1|)]$ relative to the original cell. By limiting ourselves to strain mismatch between the FeCl$_3$ superlattice and the graphene superlattice to less than 1.5%, we are able to construct heterostructure cells approximately every 5° from 0° to 30°. For angles between 30° and 60°, the graphene layer is flipped by mirror symmetry and therefore will be the same as one the of previous angles (*i.e.*, n° and (60-n)° are the same). In Table S1, we list the supercell sizes, angles, and strain mismatch for the supercells considered in this work.

| Angle | FeCl3 supercell | Graphene supercell | Strain mismatch |
|---|---|---|---|
| 0.00° | 2 × 2 | (5,0) × (0,5) | -1.41% |
| 5.82° | 7 × 7 | (18,2) × (−2,16) | 0.98% |
| 10.16° | 6 × 6 | (16,3) × (−3,13) | 0.41% |
| 15.29° | 4 × 4 | (11,3) × (−3,8) | 0.12% |
| 20.63° | 5 × 5 | (14,5) × (−5,9) | 0.31% |
| 24.18° | 6 × 6 | (17,7) × (−7,10) | -0.05% |
| 30.00° | 7 × 7 | (20,10) × (−10,10) | -0.37% |

Table S1. Mismatch angle, supercell size, and lattice percent mismatch for construction of FeCl$_3$ / graphene supercells.

## III. ATOMIC STRUCTURE OF FeCl$_2$ COMPARED TO FeCl$_3$

The atomic structures of FeCl$_2$ and FeCl$_3$ are shown in Fig. S3. The ADF image in Fig. 4 of the main text [also show in Fig. S3(a)] displays good qualitative and quantitative agreement with the STEM simulations [Fig. S3(c) and S3(b)]. There should only be one layer of iron chloride since there is only one interlayer region to intercalate into BLG. However, AB-stacked FeCl$_3$ looks quantitatively similar to FeCl$_2$. Fortunately, these two structures can be differentiated by the apparent lattice parameter in the ADF image. The lattice parameter for FeCl$_2$ is 0.347nm, and the lattice parameter for AB FeCl$_3$ is 0.701nm. The measured lattice parameter of 0.350±0.005nm is clear evidence that the other in Fig. S3(d) lattice is indeed FeCl$_2$.

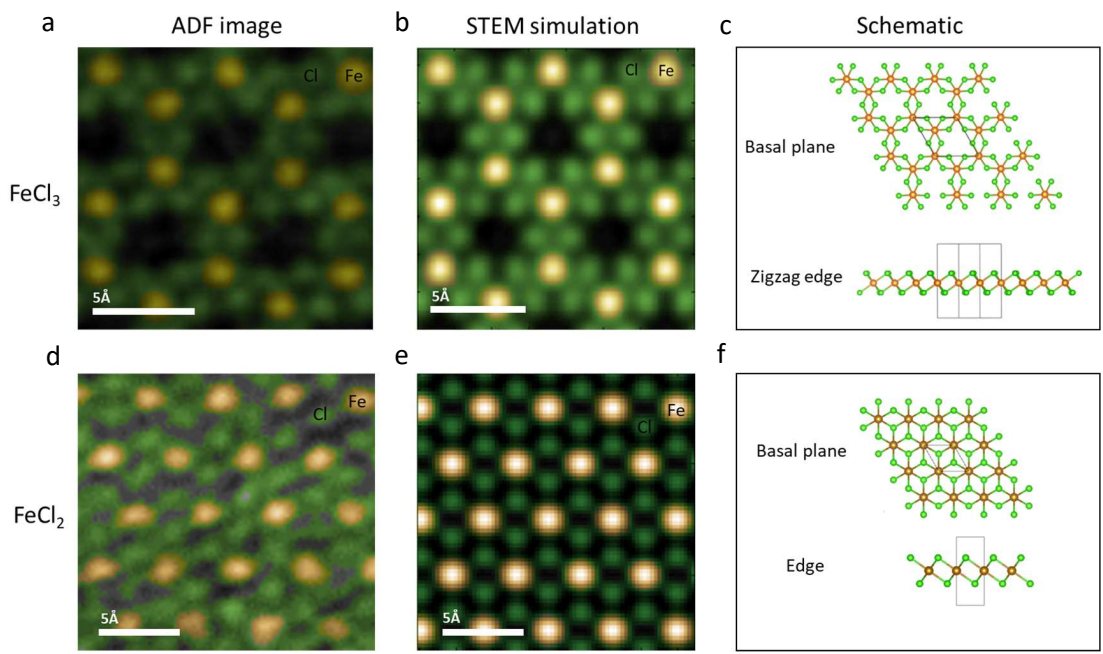

FIG. S3: (a) ADF image and (b) STEM simulation of monolayer FeCl$_3$ intercalated in BLG (c) Schematic of monolayer FeCl$_3$ (d) ADF image and (e) STEM simulation of monolayer FeCl$_3$ intercalated in BLG (f) Schematic of monolayer FeCl$_2$.

## IV. SECOND DERIVATIVE SPECTROSCPY ANALYSIS

Since multiple effects contribute to the EELS background – in particular the zero-loss peak and core-loss edges - fitting the white-line peaks in the raw data is difficult. However, fitting these peaks is critical to calculating the iron L$_3$/L$_2$ white-line intensity ratio, which in turn determines the oxidation state of the iron atoms. Analyzing the second derivative of the data allows us to determine the ratio without having to fit and subtract the background [6]. For Gaussian and Lorentzian peak shapes, the intensity ratio is equal to the ratio of the second derivative minimum multiplied by the ratio of the peak variances, which can be determined from the zeros of the second derivative. The derivation for this is shown below.

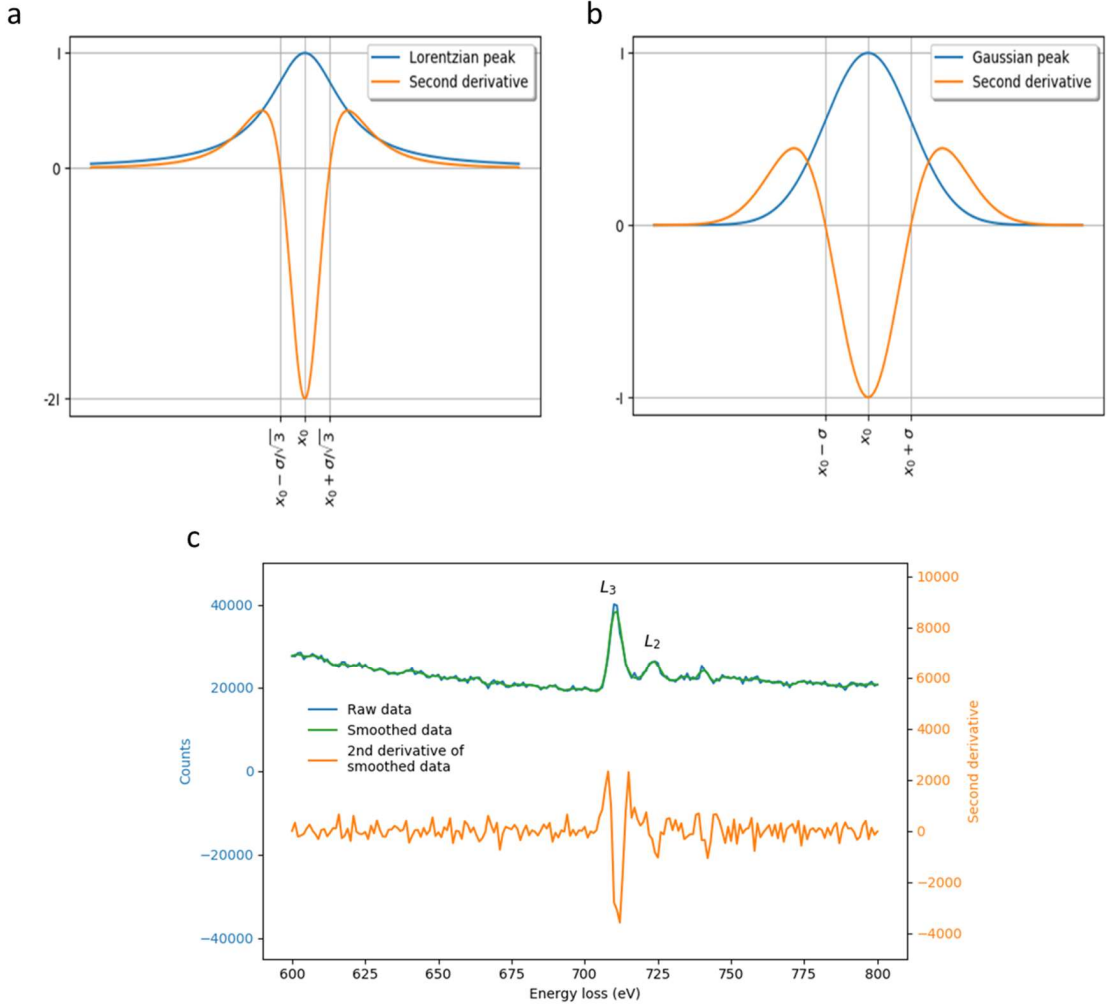

FIG. S4: A plot of a (a) Lorentzian and (b) Gaussian spectral line shape and the second derivatives. The vertical and horizontal lines in the plots show the extrema and zeros. (c) EELS of $FeCl_2$ and the second derivative of the data

Fig. S4 shows spectral lines and their second derivatives for both Lorentzian and Gaussian line shapes, as well as the EELS data from our sample that exhibits $FeCl_2$ intercalated in BLG.

For both Lorentzian and Gaussian line shapes, the $L_3/L_2$ white-line intensity is given by

$$\frac{y(x_0)_{L_3}}{y(x_0)_{L_2}} = \frac{y''(x_0)_{L_3}}{y''(x_0)_{L_2}} \frac{\alpha_{L_3}^2}{\alpha_{L_2}^2} \qquad (3)$$

where $y(x_0)$ is the maximum value at the center of the spectral line, $y''(x_0)$ is the minimum value of the second derivative, and $\alpha$ is the zero of the second derivative measured with respect to $x_0$ [i.e. $\alpha = x(y''=0)-x_0$].

## A. Derivation for Lorentzian line shape

The Lorentzian line shape is given by the function

$$L(x) = \frac{L_{max}}{1+\left(\frac{x-x_0}{\sigma}\right)^2} \tag{4}$$

where $L_{max}$ is the maximum of the spectral line, $x_0$ is the location of the spectral line, and $\sigma$ is the half width at half maximum.

The second derivative evaluated at $x_0$ is

$$L''(x_0) = -2\frac{L_{max}}{\sigma^2} \tag{5}$$

Using equation 4 and 5 to write a ratio of $L_3$ and $L_2$ intensities gives

$$\frac{L_{max,L_3}}{L_{max,L_2}} = \frac{L(x_0)_{L_3}}{L(x_0)_{L_2}} = \frac{L''(x_0)_{L_3}}{L''(x_0)_{L_2}} \frac{\sigma_{L_3}^2}{\sigma_{L_2}^2} \tag{6}$$

The half width at half maximum can also be written in terms of the zeros of the second derivative.

$$\alpha_{Lorentzian} = \pm\frac{\sigma}{\sqrt{3}} \tag{7}$$

Solving for $\sigma$ and inserting in equation 6 gives equation 3 for a Lorentzian line shape.

## B. Derivation for Gaussian line shape

The Gaussian line shape is given by the function

$$G(x) = G_{max} e^{-\frac{(x-x_0)^2}{2\sigma^2}} \tag{8}$$

where $G_{max}$ is the maximum of the spectral line, $x_0$ is the location of the spectral line, and $\sigma$ is the standard deviation.

The second derivative evaluated at $x_0$ is

$$G''(x_0) = -\frac{L_{max}}{\sigma^2} \tag{9}$$

Using equation 8 and 9 to write a ratio of $L_3$ and $L_2$ intensities gives

$$\frac{G_{\max,L_3}}{G_{\max,L_2}} = \frac{G(x_0)_{L_3}}{G(x_0)_{L_2}} = \frac{G''(x_0)_{L_3}}{G''(x_0)_{L_2}} \frac{\sigma_{L_3}^2}{\sigma_{L_2}^2} \qquad (10)$$

The half width half maximum can also be written in terms of the zeros of the second derivative.

$$\alpha_{Lorentzian} = \pm\sigma \qquad (11)$$

Solving for σ and inserting in equation 10 gives equation 3 for a Gaussian.

Fig. S4(c) displays the EELS iron edge of the sample shown in Fig. 3(c) of the main text. The experimental $L_3/L_2$ white-line intensity ratio for this section of our sample is 4±1, supporting the hypothesis that this structure is indeed $FeCl_2$.

## V. CHARGE TRANSFER IN THE BILAYER GRAPHENE – INTERCALANT SYSTEM

Using DFT, the calculated density of states for an $FeCl_3$-intercalated BLG is shown in Fig. S5(a). It exhibits a Fermi energy of 0.66 eV below the Dirac point of the graphene layers, indicating a hole carrier density of $3.2 \times 10^{13}$ cm$^{-2}$. The lower theoretical value of the relative Fermi energy compared to the experimental value is likely due to the underestimate of the Fermi velocity in the local density approximation [7].

The appearance of free carriers in graphene is typically interpreted to indicate charge transfer to or from graphene, in this case from the graphene to the intercalant. Such a conclusion, however, based solely on the shift of the Fermi level relative to the Dirac point does not take into consideration what happens projected DOS throughout the valence bands and must be corroborated by a demonstration that physical charge actually shifts in space from the graphene to the intercalant layer. Assigning charge as belonging to individual atoms by examining the charge in spheres of chosen radii or some other polyhedrons centered around individual atoms, as in Mulliken [8], Hirschfeld [9], or Bader [10] charges, entail diverse assumptions and are subject to ambiguities [11]. For the issue at hand here, however, there is a particularly robust way to examine the question of charge transfer [12]. We examine the DFT- calculated charge distribution around carbon atoms in pristine, freestanding, monolayer graphene and around several nonequivalent carbon atoms in the graphene layers of a $FeCl_3$-intercalated BLG structure. Using the nucleus of each carbon as the origin, we integrate the DFT-calculated electron density in concentric spheres of increasing radius $R$ and plot the integrated charge as a function of $R$ in Fig. S5(b). It is clear that the charge distribution around carbon atoms is indistinguishable, demonstrating that, despite the presence of free holes, the electron density in graphene remains essentially undisturbed, i.e., the graphene sheets remain charge neutral. This conclusion is corroborated by similar plots for Fe atoms in freestanding $FeCl_3$ and $FeCl_3$-intercalated BLG [Fig. S5(c)]. Once more the charge around Fe atoms remains undisturbed, demonstrating that the $FeCl_3$ layer does not actually hold any excess charge. The explanation of these results is that the wave functions of electrons nominally belonging to intercalant atoms extend over the graphene monolayers, restoring the overall charge neutrality of all carbon atoms. In other words, the manifestation of holes constitutes a redistribution of electrons in energy space, but not in physical space.

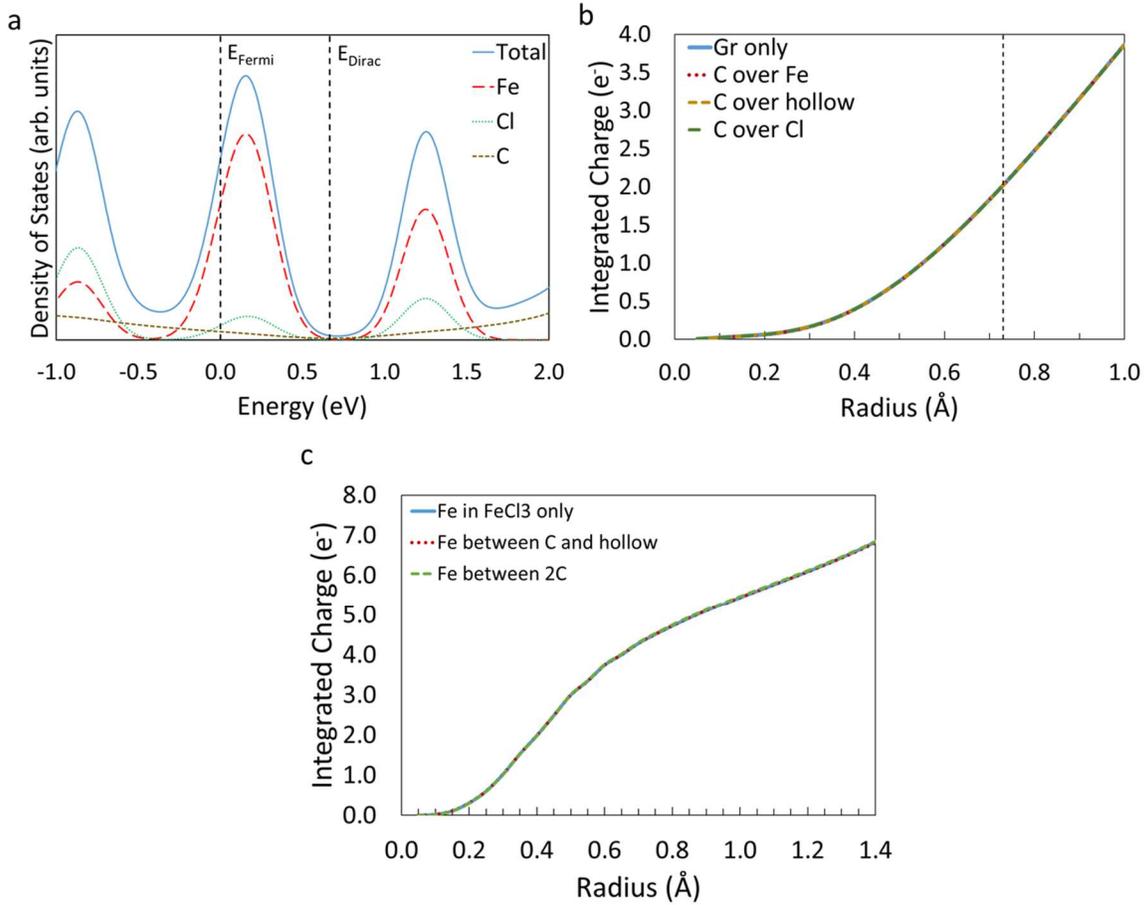

FIG. S5: (a) Total density of states and site-projected density of states for each atomic species in the $FeCl_3$-intercalated bilayer graphene system. Dashed lines represent the relative Fermi energy and Dirac-point energies, respectively. (b) Integrated charge within a sphere of radius $R$ centered about a single carbon atom as a function of $R$ in pure graphene and various positions in the $FeCl_3$-intercalated bilayer graphene system. The dashed line represents the carbon-carbon distance in graphene. (c) Integrated charge in spheres of radius $R$ centered on a single iron atom as a function of $R$ in free monolayer $FeCl_3$ and for various positions in the $FeCl_3$-intercalated bilayer graphene system.

The above conclusions regarding the absence of net physical-charge transfer accompanying the appearance of free carriers (doping) in graphene is reinforced by a simpler example, namely Li adatoms on a freestanding graphene monolayer, which are known to induce the appearance of free electrons above the Dirac point [n-type doping, Fig. S6(a), for a coverage of $8.04 \times 10^{13}$ Li/cm$^2$]. Fig. S6(b) shows the projected DOS on an adsorbed Li atom, revealing clearly that, at the chosen low concentration of Li adatoms, we have an empty Li 2s state above the Fermi level, which would immediately be interpreted as a net charge transfer, corresponding to the electrons above the Dirac point in Fig. S6(a), leaving behind Li$^+$ ions. If, however, we use the Li nucleus as the origin and integrate the charge in spheres of increasing radius R around it, we obtain a distribution that looks identical to the physical charge around a Li atom in Li metal, calculated in the same way [Fig. S6(c)], i.e., we do not have a Li$^+$ ion but neutral Li. Very similar results are obtained for a high Li density. Note that the rise that occurs in Li adatom charge above ~1 Å simply reflects the electrons from neighboring C atoms. These results are fully consistent with contour plots of the charge around Li atoms adsorbed on graphene given in Ref. [13], though those authors interpreted their results in terms of charge transfer to the region between the Li adatom and graphene C atoms.

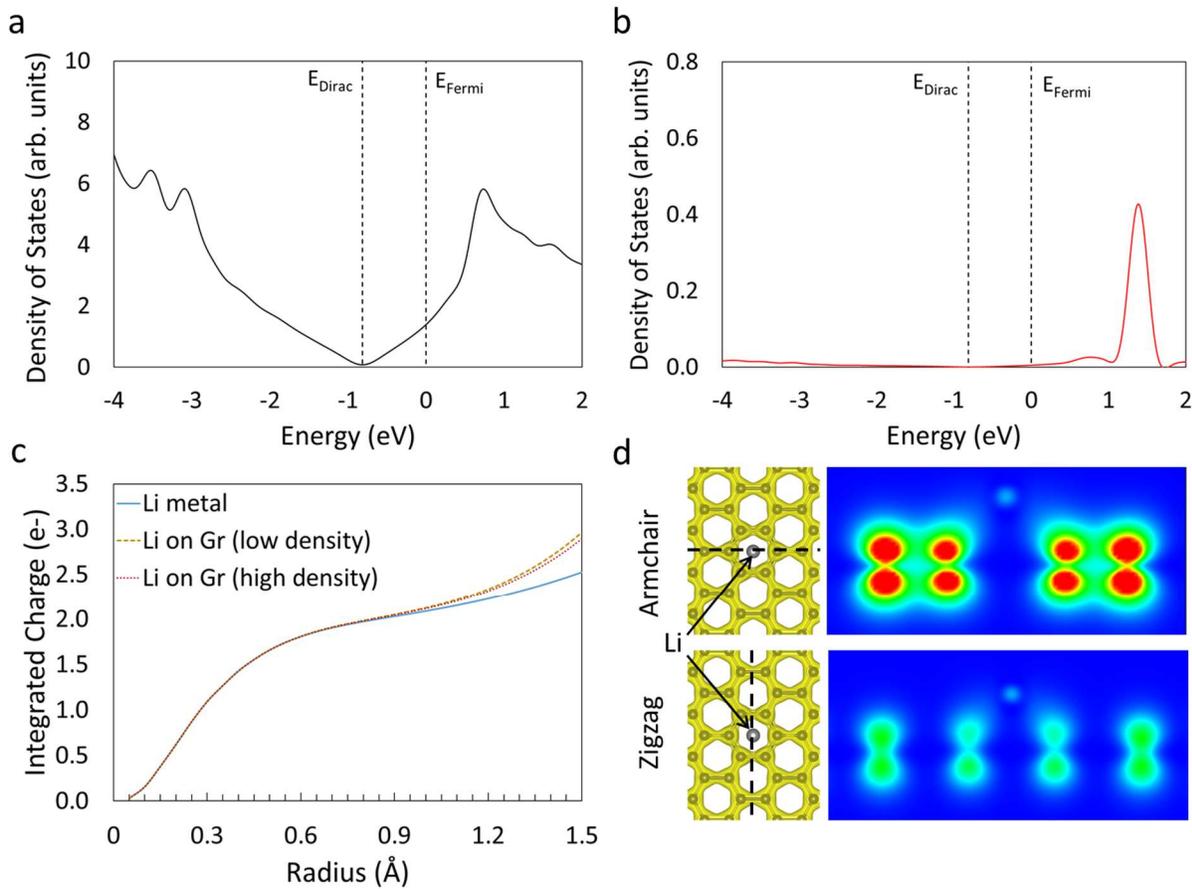

FIG. S6: The density of states projected onto (a) the carbon atoms of the graphene layer and (b) Li atoms for $8.04 \times 10^{13}$ Li/cm$^2$ concentration of Li adsorbed on graphene. (c) Comparison of the integrated charge on spheres of radius R centered on a Li atom in Li metal and the Li adsorbed on graphene in two different concentrations, as functions of $R$. (d) Two-dimensional cuts through the spatial partial charge density from the bottom of the valence band to the Fermi energy along (top) the armchair direction, cutting along carbon-carbon bonds, and (bottom) the zigzag direction, cutting between carbon-carbon bonds. Both directions show the presence of electron charge around the lithium site and distortions in the adjacent carbon $p_z$ orbitals.

Further corroboration of the above analysis is provided by the long "tail" in the DOS of Fig. S6(b) which is the result of carbon wavefunctions in the valence-band region extending over the Li atoms and restoring their neutrality. The number of states in that tail is in fact equal to the number of electrons in the sphere used to carry out the projected DOS calculations. In other words, although the Li 2s level lies above the Fermi energy and is empty, tails of carbon orbitals enter the Li space and make it neutral. Additionally, Fig. S6(d) displays two-dimensional cutouts of the electron densities in the graphene valence bands that clearly show a finite electron density around the Li atoms and distortion of the carbon orbitals. This observation indicates that, although electrons from the Li 2s state are effectively transferred to the graphene layer, the electrons from the carbon $p_z$ states extend to the Li atoms to maintain charge neutrality. These results are consistent with earlier findings that all atoms in so-called ionic crystals are effectively neutral, independent of the oxidation state [12]. The ultimate conclusion is that when adatoms or another material are in proximity to graphene, electrons in graphene are redistributed in energy space, resulting in free carriers either above or below the Dirac point (n-type or p-type doping). However, the electrons are not redistributed in physical space, resulting effectively in no charge transfer.

# VI. COMPARING STEM SIMULATIONS FOR MULTILAYER FeCl3

To determine the stacking configuration of the aligned FeCl3, STEM simulations of AB, ABA, and ABC stacked FeCl3 [Fig. S7] were compared to the ADF image [see Fig. 4(a) in the main text]. Fig. S7 demonstrates that only the simulation of ABC stacked FeCl3 with the appropriate detection angles matches the ADF image shown in the main text [Fig. 5(a)].

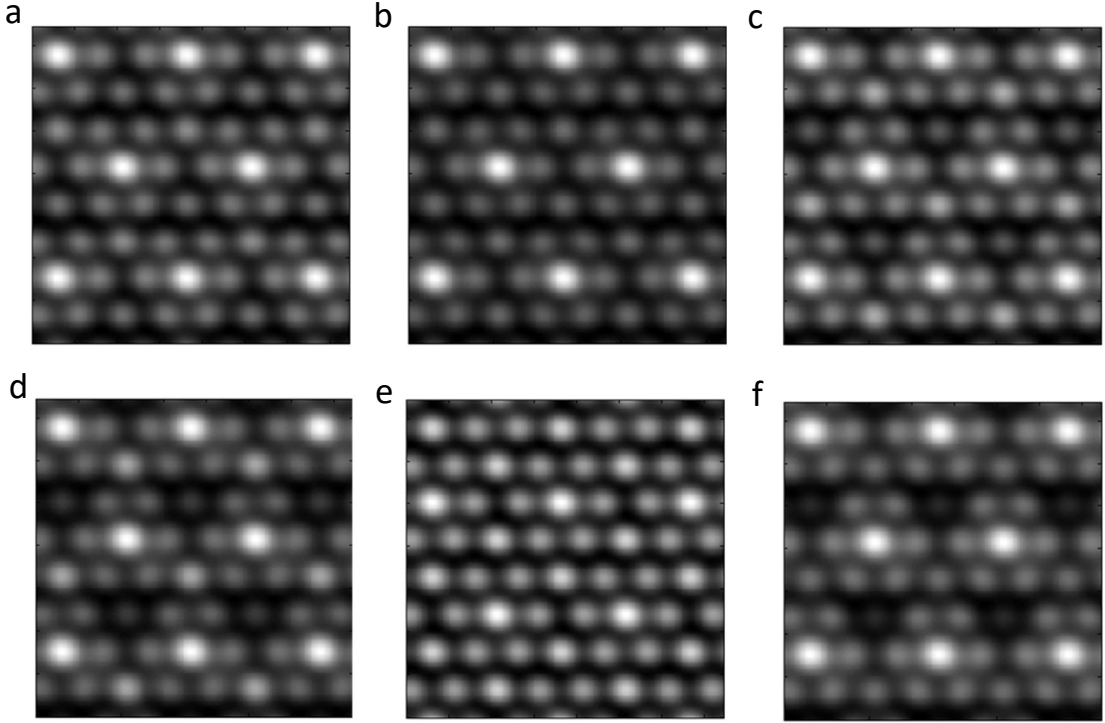

FIG. S7: STEM simulations of (a) AB (b) ABA (c) ABC stacked $FeCl_3$ with 54-200mrad detection range and (d) AB (e) ABA (f) ABC stacked $FeCl_3$ with 86-200mrad detection range

# VII. MOIRÉ LATTICE PARAMETER DERIVATION

The moiré lattice is formed by interference between two slightly mismatched lattices. One example of this concept is the formation of beats between two oscillating functions:

$$e^{i\vec{k}_1\cdot\vec{r}} + e^{i\vec{k}_2\cdot\vec{r}} = e^{i\frac{(\vec{k}_1+\vec{k}_2)}{2}\cdot\vec{r}} e^{i\frac{(\vec{k}_1-\vec{k}_2)}{2}\cdot\vec{r}} + e^{i\frac{(\vec{k}_1+\vec{k}_2)}{2}\cdot\vec{r}} e^{-i\frac{(\vec{k}_1-\vec{k}_2)}{2}\cdot\vec{r}}$$

$$= 2\left[\cos\left(\frac{\vec{k}_1+\vec{k}_2}{2}\cdot\vec{r}\right)\cos\left(\frac{\vec{k}_1-\vec{k}_2}{2}\cdot\vec{r}\right) + i\sin\left(\frac{\vec{k}_1+\vec{k}_2}{2}\cdot\vec{r}\right)\cos\left(\frac{\vec{k}_1-\vec{k}_2}{2}\cdot\vec{r}\right)\right] \quad (12)$$

Plotting the real part in one dimension gives:

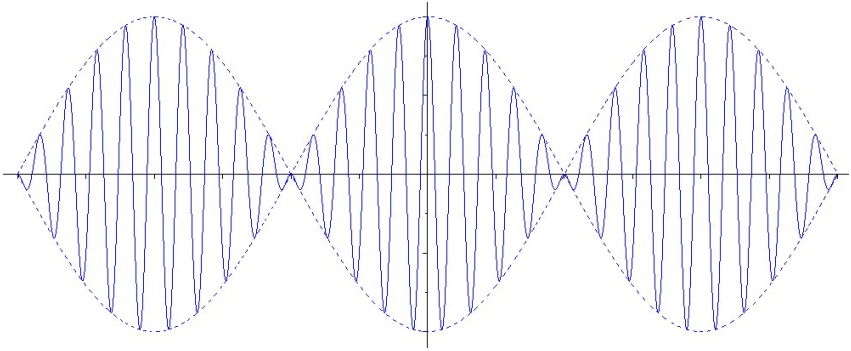

The dotted line represents the envelope function, which displays the moiré pattern, and this envelope function has a wavevector of $\vec{k_1} - \vec{k_2}$. Additionally, this derivation can be applied to any periodic function using a Fourier expansion. Thus, the reciprocal lattice vector for the moiré lattice is the difference between the reciprocal lattice vectors of the lattices that interfere to form the moiré lattice.

$$\vec{b}_{moiré} = \vec{b}_\alpha - \vec{b}_\beta \quad (13)$$

The vector $\vec{b}_{moiré}$ is the reciprocal lattice vector for the moiré lattice, and the vectors $\vec{b_\alpha}$ and $\vec{b_\beta}$ are the reciprocal lattice vectors for the α and β lattices that form the moiré lattice.

For this work, the α and β lattices are identical with a lattice parameter a=0.61nm and rotated by an angle (θ) relative to one another. Using a coordinate system with $\vec{b_\alpha}$ in the x direction:

$$\left|\vec{b}_\alpha\right| = \left|\vec{b}_\beta\right| = \frac{2\pi}{a} \quad (14)$$

$$\vec{b}_\alpha = \frac{2\pi}{a}\begin{pmatrix}1\\0\end{pmatrix} \quad (15)$$

$$\vec{b}_\beta = \begin{pmatrix}\cos\theta & -\sin\theta\\ \sin\theta & \cos\theta\end{pmatrix}\vec{b}_\alpha = \frac{2\pi}{a}\begin{pmatrix}\cos\theta\\ \sin\theta\end{pmatrix} \quad (16)$$

Combining Equations (4)-(7) yields:

$$\left|\vec{b}_{moiré}\right| = \frac{2\pi}{a}\sqrt{2(1-\cos\theta)} = \frac{2\pi}{a_{moiré}} \quad (17)$$

Solving for $a_{moiré}$ gives Equation 1 in the main text:

$$a_{moiré} = \frac{a}{\sqrt{2(1-\cos\theta)}} \quad (18)$$